\documentclass[10pt,aps,prb,amsmath,tightenlines,notitlepage]{revtex4-1}
\usepackage[latin9]{inputenc}
\usepackage{array}
\usepackage{multirow}
\usepackage{amsmath}

\makeatletter

\providecommand{\tabularnewline}{\\}

\usepackage{xcolor}
\usepackage{multirow}
\usepackage{textcomp}

\makeatother

\begin{document}
\title{Efficient evaluation of two-center Gaussian integrals in periodic
systems}
\author{Sandeep Sharma}
\email{sanshar@gmail.com}

\affiliation{Department of Chemistry, University of Colorado, Boulder, CO 80302,
USA}
\author{Gregory Beylkin}
\email{beylkin@colorado.edu}

\affiliation{Department of Applied Mathematics, University of Colorado, Boulder,
CO 80309, USA}
\begin{abstract}
By using Poisson's summation formula, we calculate periodic integrals
over Gaussian basis functions by partitioning the lattice summations
between the real and reciprocal space, where both sums converge exponentially
fast with a large exponent. We demonstrate that the summation can
be performed efficiently to calculate 2-center Gaussian integrals
over various kernels including overlap, kinetic, and Coulomb. The
summation in real space is performed using an efficient flavor of
the McMurchie-Davidson Recurrence Relation (MDRR). The expressions
for performing summation in the reciprocal space are also derived
and implemented. The algorithm for reciprocal space summation allows
us to reuse several terms and leads to significant improvement in
efficiency when highly contracted basis functions with large exponents
are used. We find that the resulting algorithm is only between a factor
of 5 to 15 slower than that for molecular integrals, indicating the
very small number of terms needed in both the real and reciprocal
space summations. An outline of the algorithm for calculating 3-center
Coulomb integrals is also provided.
\end{abstract}
\maketitle

\section{Introduction}

Electronic structure calculations of molecular systems often use the
Gaussian basis sets since they allow fast evaluation of the kinetic,
overlap, and Coulomb integrals. Further, they offer a compact representation
of the wavefunctions both in mean-field and correlated calculations.
For these reasons there has been an increasing interest in using the
Gaussian basis sets for calculations on periodic systems \citep{Guidon2009,Vandevondele2005,azarski2016,Irmler2018,Burow2009,Sun2017}.
In one popular approach \citep{pulay,pulay2,parinello} the Gaussians
are first partitioned into diffuse and sharp functions. The integrals
over diffuse Gaussians are effectively calculated using the fast Fourier
transform (FFT) whereby plane wave are used to fit the product of
two Gaussians, whereas the integrals over sharp Gaussians are either
avoided all together by the use of projected augomented wave (PAW)\citep{paw}
or are calculated explicitly. The latter approach is implemented in
the PySCF software package, an algorithm that utilizes a mixture of
Gaussians basis functions and plane waves to evaluate the integrals
using the density fitting procedure \citep{Sun2017}. However, the
cost of the integral evaluation remains high, especially if one uses
standard basis sets containing exponents that spans several orders
of magnitude typical in molecular calculations.

For periodic systems, the plane waves are the most commonly used basis
sets because \textit{inter alia} they automatically honor the translation
symmetry of the system so that various integrals, including the Coulomb
integrals, are trivially evaluated. However, plane waves do not yield
a particularly compact representations. For example, several thousand
plane wave are routinely needed and it is difficult if not impossible
to treat core electrons without the use of pseudo-potentials or projected
augmented waves. Both of these shortcomings can be overcome by using
Gaussian basis sets \citep{pisani2012hartree,doi:10.1063/1.3595514,paier2009accurate,delben13,booth2016plane,mcclain2017gaussian,Wilhelm16}.
However, one of the major bottlenecks that has prevented a more widespread
use of Gaussian basis sets for periodic systems is the cost of calculating
the Coulomb integrals that is significantly more expensive than that
for molecules. In fact many Gaussian-based mean-field calculations
avoid using such bases all together \citep{Tymczak2005,Challacombe1997,Kudin1998,Kudin2004}.

In this paper, we present an efficient algorithm for evaluation of
the 2-center integrals,
\begin{align}
\left\langle a|K|b\right\rangle = & \int\int\phi_{a}\left(\mathbf{r_{1}}\right)K\left(r_{12}\right)\phi_{b}\left(\mathbf{r_{2}}\right)d\mathbf{r}_{1}d\mathbf{r}_{2},\label{int-def}
\end{align}
where the kernel $K(r_{12})$ is a function of the inter-electron
distance $r_{12}=\left\Vert \mathbf{r}_{1}-\mathbf{r}_{2}\right\Vert $
and the associated lattice sum

\begin{eqnarray}
\left\langle a|K|b\right\rangle ^{\mbox{per}} & = & \sum_{\mathbf{P}}\int\int\phi_{a}\left(\mathbf{r_{1}}\right)K\left(\left\Vert \mathbf{r}_{1}-\mathbf{r}_{2}+\mathbf{P}\right\Vert \right)\phi_{b}\left(\mathbf{r_{2}}\right)d\mathbf{r}_{1}d\mathbf{r}_{2}\label{eq:sum over kernel-1}
\end{eqnarray}
where $\mathbf{P}=n_{1}\mathbf{a}_{1}+n_{2}\mathbf{a}_{2}+n_{3}\mathbf{a}_{3}$
are the lattice translations vectors. The kernel $K$ can be the Coulomb
kernel $1/r_{12}$, the attenuated Coulomb kernel $\mathrm{erf}(r_{12})/r_{12}$,
the Laplace kernel $\nabla_{1}\delta_{12}$ for kinetic energy operator
or the overlap kernel $\delta_{12}$, where $\delta_{12}=\delta\left(\mathbf{r}_{1}-\mathbf{r}_{2}\right)$.
The algorithm can be extended to evaluate other integrals including
those involving derivative operators (e.g. nuclear gradients) or multipole
operators. Extensions of the current algorithm for 3-center integrals
is discussed towards the end of the paper.

The current algorithm is built on top of a particularly efficient
version of the McMurchie-Davidson Recurrence Relation (MDRR) scheme
proposed recently for calculating the two center integrals \citep{Peels2020}.
We begin by presenting the salient features of this algorithm and
then show how it can be extended to treat periodic integrals. As in
most approaches, we calculate periodic integrals over Gaussian basis
functions by partitioning the lattice summations between the real
and reciprocal space by making use of the Poisson's summation formula.
A similar approach has been used in the past by Welhelm et al.\citep{Wilhelm16}.

\section{Gaussian integrals for molecular systems\label{sec:mol}}

The Gaussian type basis sets used in calculations include a pure Gaussian
multiplied by solid harmonics, a set of $2l_{a}+1$ basis function
of the form 
\[
\phi_{a}\left(\mathbf{r}\right)=S_{l_{a}}^{m_{a}}\left(\mathbf{r}-\mathbf{A}\right)\exp\left(-a\left\Vert \mathbf{r}-\mathbf{A}\right\Vert {}^{2}\right)
\]
where $m_{a}$ takes the $2l_{a}+1$ values from $-l_{a},\cdots,l_{a}$.
The solid harmonics are given in terms of homogeneous monomials,
\begin{equation}
S_{l_{a}}^{m_{a}}(\mathbf{r}-\mathbf{A})=\sum_{{a_{x},a_{y},a_{z}\atop a_{x}+a_{y}+a_{z}=l_{a}}}S_{a_{x}a_{y}a_{z}}^{l_{a}m_{a}}\left(x-A_{x}\right){}^{a_{x}}\left(y-A_{y}\right){}^{a_{y}}\left(z-A_{z}\right){}^{a_{z}},\label{eq:sph}
\end{equation}
where the summation is such that $a_{x}+a_{y}+a_{z}=l_{a}$.

A useful relation (known as Hobson's theorem \citep{dorothea,Giese}),
\begin{equation}
\left(\frac{1}{2a}\right)^{l_{a}}S_{l_{a}}^{m_{a}}\left(\nabla_{A}\right)\exp\left(-a\left\Vert \mathbf{r}-\mathbf{A}\right\Vert {}^{2}\right)=S_{l_{a}}^{m_{a}}\left(\mathbf{r}-\mathbf{A}\right)\exp\left(-a\left\Vert \mathbf{r}-\mathbf{A}\right\Vert {}^{2}\right),\label{eq:hobson}
\end{equation}
shows that a Gaussian basis function with non-zero $l_{a}$ can be
obtained by differentiating a simple Gaussian an appropriate number
of times with respect to its center. Using this relation, we can rewrite
Equation~\ref{int-def} as 
\begin{equation}
\left\langle a|K|b\right\rangle =\left(\frac{1}{2a}\right)^{l_{a}}S_{l_{a}}^{m_{a}}\left(\nabla_{A}\right)\left(\frac{1}{2b}\right)^{l_{b}}S_{l_{b}}^{m_{b}}\left(\nabla_{B}\right)I_{0}\left(a,b,T\right),\label{eq:basic}
\end{equation}
\begin{align}
I_{0}\left(a,b,T\right)= & \int\exp\left(-a\left\Vert \mathbf{r}_{1}-\mathbf{A}\right\Vert {}^{2}\right)K\left(r_{12}\right)\exp\left(-b\left\Vert \mathbf{r}_{2}-\mathbf{B}\right\Vert {}^{2}\right)d\mathbf{r}_{1}d\mathbf{r}_{2}\\
= & \left(\frac{\pi}{a+b}\right)^{3/2}\int e^{-\rho\left\Vert \mathbf{A-B-r}\right\Vert {}^{2}}K\left(r\right)d\mathbf{r}\\
= & \left(\frac{\pi}{a+b}\right)^{3/2}2\pi e^{-\rho\left\Vert \mathbf{A}-\mathbf{B}\right\Vert {}^{2}}\int_{0}^{\infty}\int_{0}^{\pi}e^{-\rho r^{2}}e^{2\rho\mathbf{\left\Vert A-B\right\Vert }r\cos\theta}K\left(r\right)r^{2}dr\sin\theta d\theta\\
= & \left(\frac{\pi}{a+b}\right)^{3/2}2\pi e^{-\rho\left\Vert \mathbf{A}-\mathbf{B}\right\Vert {}^{2}}\int_{0}^{\infty}e^{-\rho r^{2}}\frac{r\sinh\left(2\rho\mathbf{\left\Vert A-B\right\Vert }r\right)}{\rho\mathbf{\left\Vert A-B\right\Vert }}K\left(r\right)dr\\
= & \left(\frac{\pi}{a+b}\right)^{3/2}G_{0}\left(\rho,T\right)
\end{align}
where $r=\left\Vert \mathbf{r}\right\Vert =\left\Vert \mathbf{r}_{1}-\mathbf{r}_{2}\right\Vert $,
$\rho=\frac{ab}{a+b}$, $T=\rho\left\Vert \mathbf{A}-\mathbf{B}\right\Vert {}^{2}$
and $G_{0}\left(\rho,T\right)$ ends up being a 1-D integral that
can be evaluated analytically for several commonly used kernels\citep{ahlrichs04,ahlrichs06}.

It is expensive to directly evaluate all $(2l_{a}+1)(2l_{b}+1)$ derivatives
of Equation~\ref{eq:basic} to compute the integrals. Instead we
can use the MDRR motivated by the observation that 
\begin{align}
S_{l_{a}}^{m_{a}}\left(\nabla_{A}\right)= & S_{l_{a}}^{m_{a}}\left(\nabla_{R}\right)\nonumber \\
S_{l_{a}}^{m_{a}}\left(\nabla_{B}\right)= & -S_{l_{a}}^{m_{a}}\left(\nabla_{R}\right)\label{eq:deriv}
\end{align}
where $\mathbf{R}=\mathbf{A-B}$. Combining Equations~\ref{int-def},
\ref{eq:hobson}, \ref{eq:sph} and \ref{eq:deriv}, we obtain 
\begin{align}
(\left\langle a|K|b\right\rangle )= & \left(\frac{\pi}{a+b}\right)^{3/2}\left(\frac{1}{2a}\right)^{l_{a}}\left(\frac{-1}{2b}\right)^{l_{b}}\nonumber \\
 & \sum_{{a_{x}a_{y}a_{z}\atop a_{x}+a_{y}+a_{z}=l_{a}}}S_{a_{x}a_{y}a_{z}}^{l_{a}m_{a}}\sum_{{b_{x}b_{y}b_{z}\atop b_{x}+b_{y}+b_{z}=l_{b}}}S_{b_{x}b_{y}b_{z}}^{l_{b}m_{b}}\left[\mathbf{c}\right]{}^{(0)},\label{eq:solidInts}
\end{align}
where $\mathbf{c=a+b}$ and we have used the shorthand notation following
\citep{gill89,gill94},
\begin{align}
\left[\mathbf{c}\right]{}^{(m)}= & \left(2\rho\right){}^{m}\frac{\partial^{c_{x}}}{\partial R_{x}^{c_{x}}}\frac{\partial^{c_{y}}}{\partial R_{y}^{c_{y}}}\frac{\partial^{c_{z}}}{\partial R_{z}^{c_{z}}}\frac{\partial^{m}}{\partial T^{m}}G_{0}\left(\rho,T\right)\\
= & \left(-2\rho\right){}^{m}\frac{\partial^{c_{x}}}{\partial R_{x}^{c_{x}}}\frac{\partial^{c_{y}}}{\partial R_{y}^{c_{y}}}\frac{\partial^{c_{z}}}{\partial R_{z}^{c_{z}}}G_{m}\left(\rho,T\right)\label{eq:gm}
\end{align}
where $G_{m}\left(\rho,T\right)=\left(-\frac{\partial}{\partial T}\right)^{m}G_{0}\left(\rho,T\right)$.
We have introduced the extra index $m$ which tells us the order of
derivative of $G_{0}\left(\rho,T\right)$ with respect to $T$. In
the final integrals we always have $m=0$ but values with non-zero
$m$ are useful intermediates that appear in the recursion of the
MDRR scheme,
\begin{align}
\left[\mathbf{c}\right]{}^{(m)}=R_{i}\left[\mathbf{c}-\mathbf{1}_{i}\right]{}^{(m+1)}+(r_{i}-1)\left[\mathbf{c}-\mathbf{2}_{i}\right]{}^{(m+1)} & ,\label{eq:recursion}
\end{align}
where $\mathbf{1}_{i}$ is a unit vector along direction $i\in\left\{ x,y,z\right\} $
and $\mathbf{2}_{i}=2\mathbf{1}_{i}$. In the MDRR scheme one begins
by evaluation of $\left[0\right]{}^{(m)}$ for all $m\leq l_{a}+l_{b}$.
Then, by using the recursion relation in Equation~\ref{eq:recursion},
we obtain the integrals $\left[\mathbf{c}\right]{}^{(0)}$ for all
possible vectors $\mathbf{c}=\{c_{x},c_{y},c_{z}\}$ such that $c_{x}+c_{y}+c_{z}=l_{a}+l_{b}$.
These $\left[\mathbf{c}\right]{}^{(0)}$ can then be inserted in Equation~\ref{eq:solidInts}
along with the known coefficients $S_{a_{x}a_{y}a_{z}}^{lm}$ to evaluate
the desired integrals.

As a result, the calculation of molecular 2-electron integrals involves
the following four steps:
\begin{enumerate}
\item Calculate the $2(l_{a}+l_{b})+1$ quantities $\left[\mathbf{0}\right]{}^{m}=\left(-2\rho\right){}^{m}G_{m}\left(\rho,T\right)$,
where $m$ can take any value from $-l_{a}-l_{b},\cdots,l_{a}+l_{b}$.
As an example, expression for $G_{m}\left(\rho,T\right)$ are tabulated
in Table~\ref{tab:G0} for several kernels where
\[
F_{m}\left(T\right)=\left(-\frac{\partial}{\partial T}\right)^{m}F_{0}\left(T\right)
\]
and 
\[
F_{0}\left(T\right)=\frac{1}{2}\sqrt{\frac{\pi}{T}}\mbox{erf}\left(\sqrt{T}\right)
\]
is the Boys function.
\item We then contract these primitive quantities using the contraction
coefficients to obtain the contracted $2(l_{a}+l_{b})+1$ quantities
$\left[\mathbf{0}\right]{}^{m}$.
\item Use MDRR scheme to evaluate $\left(2l_{a}+1\right)\left(2l_{b}+1\right)$
integrals $\left[\mathbf{c}\right]{}^{(0)}$ where $c_{x}+c_{y}+c_{z}=l_{a}+l_{b}$.
\item Use the solid harmonics coefficients to generate the desired integral
$\left\langle a|K|b\right\rangle $ from $\left[\mathbf{c}\right]{}^{(0)}$.
\end{enumerate}

\section{Gaussian integrals for periodic systems}

Having outlined the MDRR scheme to calculate the molecular integrals,
we now describe modifications needed to generate integrals over periodic
systems. In periodic systems the integrals of interest are 
\begin{eqnarray}
\left\langle a|K|b\right\rangle ^{\mbox{per}} & = & \sum_{\mathbf{P}}\int\int\phi_{a}\left(\mathbf{r_{1}}\right)K\left(\left\Vert \mathbf{r}_{1}-\mathbf{r}_{2}+\mathbf{P}\right\Vert \right)\phi_{b}\left(\mathbf{r_{2}}\right)d\mathbf{r}_{1}d\mathbf{r}_{2}\label{eq:sum over kernel}\\
 & = & \sum_{\mathbf{P}}\int\int\phi_{a}\left(\mathbf{r_{1}}\right)K\left(\left\Vert \mathbf{r}_{1}-\mathbf{r}_{2}\right\Vert \right)\phi_{b}\left(\mathbf{r_{2}}+\mathbf{P}\right)d\mathbf{r}_{1}d\mathbf{r}_{2}\nonumber \\
 & = & \sum_{\mathbf{P}}\left\langle a|K|b_{\mathbf{P}}\right\rangle \label{eq:sum}
\end{eqnarray}
where $\mathbf{P}=n_{1}\mathbf{a}_{1}+n_{2}\mathbf{a}_{2}+n_{3}\mathbf{a}_{3}$
are the lattice translations vectors, $n_{i}$ are the integers, $\mathbf{a}_{i}$
are the three primitive lattice translations vectors and $|\left.b_{\mathbf{P}}\right\rangle $
represents the Gaussian type orbital $|\left.b\right\rangle $ that
has been translated by the vector $\mathbf{P}$. The equation is useful
for calculating the Gamma point integrals and conversion to integrals
with translational symmetry. Notice that in Equation~\ref{eq:solidInts}
only the quantity $\left[\mathbf{c}\right]{}^{(0)}$ depends on $\mathbf{P}$.
In particular, the periodic integral can be written as 
\begin{align}
\left\langle a|K|b\right\rangle ^{\mbox{per}}= & \left(\frac{\pi}{a+b}\right)^{3/2}\left(\frac{1}{2a}\right)^{l_{a}}\left(\frac{-1}{2b}\right)^{l_{b}}\nonumber \\
 & \sum_{{a_{x}a_{y}a_{z}\atop a_{x}+a_{y}+a_{z}=l_{a}}}S_{a_{x}a_{y}a_{z}}^{l_{a}m_{a}}\sum_{{b_{x}b_{y}b_{z}\atop b_{x}+b_{y}+b_{z}=l_{b}}}S_{b_{x}b_{y}b_{z}}^{l_{b}m_{b}}\sum_{\mathbf{P}}\left[\mathbf{c}\left(\mathbf{P}\right)\right]{}^{(0)},\label{eq:fullexp}
\end{align}
where 
\[
\left[\mathbf{c}\left(\mathbf{P}\right)\right]^{(0)}=\frac{\partial^{c_{x}}}{\partial R_{x}^{c_{x}}}\frac{\partial^{c_{y}}}{\partial R_{y}^{c_{y}}}\frac{\partial^{c_{z}}}{\partial R_{z}^{c_{z}}}G_{0}\left(\rho,T\left(\mathbf{P}\right)\right),
\]
$T\left(\mathbf{P}\right)=\rho\left\Vert \mathbf{A-B-P}\right\Vert {}^{2}$
and three Cartesian coordinates of $\mathbf{R}$ are $R_{i}=A_{i}-B_{i}-P_{i}$.
It is also useful to observe that 
\begin{align}
\sum_{\mathbf{P}}\left[\mathbf{c}\left(\mathbf{P}\right)\right]^{(0)}= & \frac{\partial^{c_{x}}}{\partial R_{x}^{c_{x}}}\frac{\partial^{c_{y}}}{\partial R_{y}^{c_{y}}}\frac{\partial^{c_{z}}}{\partial R_{z}^{c_{z}}}\sum_{\mathbf{P}}\left[\mathbf{0\left(P\right)}\right]\\
= & \frac{\partial^{c_{x}}}{\partial R_{x}^{c_{x}}}\frac{\partial^{c_{y}}}{\partial R_{y}^{c_{y}}}\frac{\partial^{c_{z}}}{\partial R_{z}^{c_{z}}}\sum_{\mathbf{P}}G_{0}\left(\rho,T\left(\mathbf{P}\right)\right),\label{eq:sumder}
\end{align}
which means that the summation and the derivatives commute. We will
make use of this property when performing summation in reciprocal
space (see Section~\ref{sec:recsum}).

These equations give a straightforward way of performing lattice summation
and the only difference from the molecular algorithm is that one performs
steps 1 and 2 (see the end of previous section) for each value of
$\mathbf{P}$ to accumulate $\left[\mathbf{c}\left(\mathbf{P}\right)\right]^{(0)}$
for increasing values of $P$ until the sum $\sum_{\mathbf{P}}\left[\mathbf{c}\left(\mathbf{P}\right)\right]^{(0)}$
converges. Finally step 3 to obtain the desired integrals is performed
only once after the results of the lattice summation are available.

\begin{table*}[t]
\centering{}%
\begin{tabular}{lcccc}
\hline 
Kernel & ~~~ & $G_{m}\left(\rho,T\right)$ & ~~~ & $\hat{G}_{0}\left(\rho,\mathbf{G}\right)$\tabularnewline
$K\left(\mathbf{r}\right)$ &  &  &  & \tabularnewline
\hline 
$\delta\left(\mathbf{r}\right)$ &  & $\exp\left(-T\right)$ &  & $\left(\frac{\pi}{\rho}\right)^{3/2}\exp\left(-i\mathbf{G}\cdot\mathbf{R}\right)\exp\left(-\frac{G^{2}}{4\rho}\right)/\Omega$\tabularnewline
$\frac{1}{r}$ &  & $\frac{2\pi}{\rho}F_{m}\left(T\right)$ &  & $\pi\left(\frac{\pi}{\rho}\right)^{3/2}\exp\left(-i\mathbf{G}\cdot\mathbf{R}\right)\exp(-\frac{G^{2}}{4\rho})/\left(4\Omega G^{2}\right)$\tabularnewline
$\frac{\mathrm{erf}\left(\omega r\right)}{r}$ &  & $\frac{2\pi\eta}{\rho}F_{m}\left(T\eta^{2}\right)$ &  & $\pi\left(\frac{\pi}{\rho}\right)^{3/2}\exp\left(-i\mathbf{G}\cdot\mathbf{R}\right)\exp(-\frac{G^{2}}{\eta^{2}4\rho})/\left(4\Omega G^{2}\right)$\tabularnewline
\hline 
\end{tabular}\caption{The table shows the expressions for $G_{m}\left(\rho,T\right)$ in
Equation~\ref{eq:gm} and $\hat{G}_{0}(\rho,\mathbf{G})$ in Equation~\ref{eq:ksum}
for commonly used kernels to evaluate Gaussian integrals. Here $\mathbf{R=A-B}$
is the vector connecting one Gaussian $|\left.a\right\rangle $ with
$\left.|b\right\rangle $, $T=\rho|\mathbf{R}|^{2}$, $\eta=\frac{\omega}{\sqrt{\omega^{2}+\rho}}$
and $\Omega$ is the volume of the unit cell.}
\label{tab:G0}
\end{table*}

\subsection{Summation in reciprocal space}

\label{sec:recsum} Although spatial summation can be carried out
for all kernels (the Coulomb kernel required special considerations
since, as written, the sum diverges; see next section) the sum converges
slowly when either $a$ or $b$ are small. A straightforward technique
for avoiding this is to perform the summation in reciprocal space
by making use of the Poisson's summation formula which states that
\begin{align}
\sum_{\mathbf{P}}f(\mathbf{P})=\sum_{\mathbf{G}}\hat{f}(\mathbf{G}) & ,
\end{align}
where $\hat{f}$ is the Fourier transform of $f$ and is given by
\begin{align}
\hat{f}(\mathbf{G})=\int\exp(-i2\pi\mathbf{G}\cdot\mathbf{P})f(\mathbf{P})d\mathbf{P}.
\end{align}
In these equations $\mathbf{G}=m_{1}\mathbf{A}_{1}+m_{2}\mathbf{A}_{2}+m_{3}\mathbf{A}_{3}$
are the reciprocal lattice vectors, $m_{i}$ are integers and $\mathbf{A}_{i}$
are the primitive lattice vectors of the reciprocal lattice such that
$\mathbf{A}_{i}\cdot\mathbf{a}_{j}=2\pi\delta_{ij}$.

Now by making use of the Poisson's summation formula in Equation~\ref{eq:sumder}
and Equation~\ref{eq:fullexp}, we obtain 
\begin{align}
\left\langle a|K|b\right\rangle {}^{per}= & \left(\frac{\pi}{a+b}\right)^{3/2}\left(\frac{1}{2a}\right)^{l_{a}}\left(\frac{-1}{2b}\right)^{l_{b}}\sum_{{a_{x}a_{y}a_{z}\atop a_{x}+a_{y}+a_{z}=l_{a}}}S_{a_{x}a_{y}a_{z}}^{l_{a}m_{a}}\sum_{{b_{x}b_{y}b_{z}\atop b_{x}+b_{y}+b_{z}=l_{b}}}S_{b_{x}b_{y}b_{z}}^{l_{b}m_{b}}\sum_{\mathbf{P}}\left[\mathbf{c\left(P\right)}\right]^{(0)}\label{eq:ksum}\\
%
= & \left(\frac{\pi}{a+b}\right)^{3/2}\left(\frac{1}{2a}\right)^{l_{a}}\left(\frac{-1}{2b}\right)^{l_{b}}\sum_{{a_{x}a_{y}a_{z}\atop a_{x}+a_{y}+a_{z}=l_{a}}}S_{a_{x}a_{y}a_{z}}^{l_{a}m_{a}}\sum_{{b_{x}b_{y}b_{z}\atop b_{x}+b_{y}+b_{z}=l_{b}}}S_{b_{x}b_{y}b_{z}}^{l_{b}m_{b}}\frac{\partial^{c_{x}}}{\partial R_{x}^{c_{x}}}\frac{\partial^{c_{y}}}{\partial R_{y}^{c_{y}}}\frac{\partial^{c_{z}}}{\partial R_{z}^{c_{z}}}\sum_{\mathbf{P}}G_{0}\left(\rho,T\mathbf{\left(P\right)}\right)\\
= & \left(\frac{\pi}{a+b}\right)^{3/2}\left(\frac{1}{2a}\right)^{l_{a}}\left(\frac{-1}{2b}\right)^{l_{b}}\sum_{{a_{x}a_{y}a_{z}\atop a_{x}+a_{y}+a_{z}=l_{a}}}S_{a_{x}a_{y}a_{z}}^{l_{a}m_{a}}\sum_{{b_{x}b_{y}b_{z}\atop b_{x}+b_{y}+b_{z}=l_{b}}}S_{b_{x}b_{y}b_{z}}^{l_{b}m_{b}}\frac{\partial^{c_{x}}}{\partial R_{x}^{c_{x}}}\frac{\partial^{c_{y}}}{\partial R_{y}^{c_{y}}}\frac{\partial^{c_{z}}}{\partial R_{z}^{c_{z}}}\sum_{\mathbf{G}}\hat{G}_{0}\left(\rho,\mathbf{G}\right).
\end{align}
Here $\hat{G}_{0}$ is the Fourier transform of $G_{0}$; the expressions
for a few kernels are given in Table~\ref{tab:G0}. Notice that when
$\rho$ is small it is advantageous to perform the summation in reciprocal
space (and vice versa). The derivatives with respect to $R_{i}$ are
easily evaluated 
\begin{align}
\sum_{\mathbf{P}}\left[\mathbf{c\left(P\right)}\right]^{(0)}= & \frac{\partial^{c_{x}}}{\partial R_{x}^{c_{x}}}\frac{\partial^{c_{y}}}{\partial R_{y}^{c_{y}}}\frac{\partial^{c_{z}}}{\partial R_{z}^{c_{z}}}\sum_{\mathbf{P}}G_{0}\left(\rho,\mathbf{P}\right)\\
= & \sum_{\mathbf{G}}\left(iG_{x}\right){}^{c_{x}}\left(iG_{y}\right){}^{c_{y}}\left(iG_{z}\right){}^{c_{z}}\hat{G}_{0}\left(\rho,\mathbf{G}\right)\label{eq:crec}
\end{align}

The algorithm for calculating the integrals by reciprocal summation
then consists of three steps:
\begin{enumerate}
\item First we calculate the primitive $\left(l_{a}+l_{b}\right)\left(l_{a}+l_{b}+1\right)$
integrals, where $c_{x}+c_{y}+c_{z}=l_{a}+l_{b}$ by summing over
the reciprocal vectors in Equation~\ref{eq:crec}.
\item These primitive integrals are now contracted to obtain the contracted
integrals.
\item These contracted integrals are then substituted in Equation~\ref{eq:ksum}
to obtain the necessary integrals containing solid harmonics.
\end{enumerate}

\subsection{Coulomb kernel}

The summation over Coulomb kernel requires special care because if
one uses the bare Coulomb kernel $1/r$ then $\sum_{\mathbf{P}}G_{0}\left(\rho,T\left(\mathbf{P}\right)\right)$
is divergent. Here we will derive the relevant expressions by starting
from the summation in the reciprocal space, $\sum_{\mathbf{G}}\hat{G}_{0}\left(\rho,\mathbf{G}\right)$
which is also divergent but only due to the $G=0$ term when $\rho\neq0$
(see the Table~\ref{tab:G0} for the expression). The standard procedure
for obtaining convergent Coulomb integrals calls for performing the
summation in reciprocal space and eliminating the $G=0$ term. This
makes the total system charge-neutral by removing a constant background
charge and imposes conducting boundary conditions which ensure that
the surface term disappears. Performing the summation entirely in
reciprocal space leads to acceptable results except for the fact that
for large $\rho$ the sum does not converge rapidly. Therefore, we
use the approach of Ewald's summation \citep{ewald} and introduce
a parameter $\eta$ which allows us to perform part of the summation
in real space and the remaining in reciprocal space so that both sums
converge rapidly. Below we derive the appropriate expression by starting
with $\hat{G}_{0}\left(\rho,\mathbf{G}\right)$ from Table~\ref{tab:G0}
(we have removed the constant prefactors to avoid clutter).
\begin{widetext}
\begin{align}
\sum_{\mathbf{G\neq0}}\hat{G}_{0}\left(\rho,\mathbf{G}\right)= & \sum_{\mathbf{G\neq0}}\exp\left(-i\mathbf{G}\cdot\mathbf{R}\right)\frac{\exp\left(-\frac{G^{2}}{4\rho}\right)}{G^{2}}\nonumber \\
= & \sum_{\mathbf{G\neq0}}\exp\left(-i\mathbf{G}\cdot\mathbf{R}\right)\left(\frac{\exp\left(-\frac{G^{2}}{4\eta^{2}\rho}\right)}{G^{2}}-\frac{\exp\left(-\frac{G^{2}}{4\eta^{2}\rho}\right)}{G^{2}}+\frac{\exp\left(-\frac{G^{2}}{4\rho}\right)}{G^{2}}\right)\nonumber \\
= & \sum_{\mathbf{G\neq0}}\exp\left(-i\mathbf{G}\cdot\mathbf{R}\right)\frac{\exp\left(-\frac{G^{2}}{4\eta^{2}\rho}\right)}{G^{2}}+\sum_{\mathbf{G}}\exp\left(-i\mathbf{G}\cdot\mathbf{R}\right)\frac{\exp\left(-\frac{G^{2}}{4\rho}\right)-\exp\left(-\frac{G^{2}}{4\eta^{2}\rho}\right)}{G^{2}}\nonumber \\
- & \lim_{\delta\rightarrow0}\frac{\exp\left(-\frac{\delta^{2}}{4\rho}\right)-\exp\left(-\frac{\delta^{2}}{4\eta^{2}\rho}\right)}{\delta^{2}}\\
= & \sum_{\mathbf{G\neq0}}\exp\left(-i\mathbf{G}\cdot\mathbf{R}\right)\frac{\exp\left(-\frac{G^{2}}{4\eta^{2}\rho}\right)}{G^{2}}+\sqrt{\frac{4\rho}{\pi^{3}}}\sum_{\mathbf{P}}F_{0}\left(T\left(\mathbf{P}\right)\right)-\eta F_{0}\left(\eta^{2}T\left(\mathbf{P}\right)\right)-\frac{1}{4\omega^{2},}\label{eq:coulSum}
\end{align}
where $G=\left\Vert \mathbf{G}\right\Vert $ and $\omega=\frac{\rho\eta}{\sqrt{1-\eta^{2}}}$
(which is the same relation between $\omega$ and $\eta$ in Table~\ref{tab:G0}).
In the second step above, we have added and subtracted the $\mathbf{G}=0$
contribution to the summation and, in the third step, we have used
the Poisson summation to convert from the reciprocal space to the
real space summation. By picking an appropriate value of $\eta$,
we can ensure the fast convergence in both real and reciprocal spaces
\citep{perram}. The real space summation follows closely the algorithm
at the end of Section~\ref{sec:mol} and the reciprocal space summation
is identical to the one in Section~\ref{sec:recsum}. The constant
background term $\frac{1}{4\omega^{2}}$ is only present for $s$-type
Gaussians because it gets eliminated when one takes derivatives to
obtain integrals for higher angular momentum Gaussians (see Equation~\ref{eq:crec}).
\end{widetext}

\subsection{Selecting screening parameter and reusing reciprocal summations}

Here we focus on the Coulomb kernel not only because it is most expensive
to evaluate but, also, because the cost of performing mean field calculations
is determined to a large extent by the efficiency with which three-/four-center
integrals over Coulomb kernel are evaluated. (While this is outside
the scope of current work, we give an outline of an algorithm in Section
\ref{sec:3-center} to be implemented in a future publication). For
calculation of the Coulomb integrals we may need to perform summations
in both the real and reciprocal space. The summation in the reciprocal
space consists of evaluating 
\begin{equation}
\sum_{\mathbf{G}\neq0}\left(iG_{x}\right){}^{c_{x}}\left(iG_{y}\right){}^{c_{y}}\left(iG_{z}\right){}^{c_{z}}\exp\left(-i\mathbf{G}\cdot\mathbf{R}\right)\frac{\exp\left(-\frac{G^{2}}{4\eta^{2}\rho}\right)}{G^{2}},\label{eq:Ksum}
\end{equation}
where $c_{x}+c_{y}+c_{z}=l_{a}+l_{b}$ and $\mathbf{R}=\mathbf{A}-\mathbf{B}$.
The sum in Equation~\ref{eq:Ksum} depends only on $\mathbf{R}$
and $\eta^{2}\rho$; for a fixed value of $\eta^{2}\rho$ there are
only as many distinct values of $\mathbf{R}$ as the number of unique
atom pairs. This allows us to reuse summations as follows:
\begin{enumerate}
\item We begin by choosing a threshold value of $\eta^{2}\rho$ which, for
the purposes of this paper, is always chosen to be $T/8$, where $T$
is the smallest of the three dimensions of the unit cell.
\item If the pair of basis functions are diffuse enough such that the value
of $\rho$ is smaller than $T/8$, then the value of the integral
is calculated entirely in the reciprocal space and no part of the
summation is done in real space, i.e. $\eta$=1 is used for these
pairs of basis functions.
\item If the pair of basis functions are sharp such that the value of $\rho$
is greater than $T/8$, then we chose $\eta=\sqrt{\frac{T}{8\rho}}$.
This ensures that the summation in the reciprocal space is given in
Equation \ref{eq:Ksum}. However, instead of performing this summation
for each pair of sharp basis functions (i.e. $O\left(N_{basis}^{2}\right)$
times, we perform this summation once at the beginning of the calculation
for each unique value of $\mathbf{R}$ of which there are only $O\left(N_{atom}^{2}\right)$
terms, usually 1 or 2 order of magnitude fewer than the total number
$O\left(N_{basis}^{2}\right)$. The real space summation is performed
as usual with the attenuated Coulomb kernel which converges quickly
(usually within less than a unit cell to numerical accuracy). Effectively,
we only need to perform summation in the real space.
\end{enumerate}
The above illustrates the fact that one only needs to perform either
real or reciprocal space summations to calculate the Coulomb integrals.
For the purposes of this paper we have always used a value of $T/8$
as the threshold of $\eta^{2}\rho$ which gives reasonable timing
for all the systems studied here, but we do find that for some calculations
shown below the cost can be reduced further by chose a threshold other
than $T/8$. This will be investigated more thoroughly in a future
publication.

\section{Results}

The algorithm outlined above has been implemented in a custom branch
of PySCF\citep{pyscf1,pyscf2}. Here we show the performance of some
typical periodic systems with large auxiliary Gaussian basis sets.
The choice of the basis is partly motivated by the fact that these
basis sets are available for elements of the entire periodic table
and we expect that high accuracy all-electron calculations can be
performed with these. The pyscf input files with geometries of the
systems are given in the supplementary information. All calculations
were performed on a single core of the Intel\textsuperscript{\textregistered}
Xeon\textsuperscript{\textregistered} CPU E5-2680 v4 @ 2.40GHz processor.

\begin{table}[htb]
\centering{}%
\begin{tabular}{ccccrrrrrrrrrrr}
\hline 
 &  &  &  & \multicolumn{11}{c}{\textbf{Time Elapsed (s)}}\tabularnewline
\cline{5-15} \cline{6-15} \cline{7-15} \cline{8-15} \cline{9-15} \cline{10-15} \cline{11-15} \cline{12-15} \cline{13-15} \cline{14-15} \cline{15-15} 
 &  &  &  & \multicolumn{3}{c}{Molecule} & ~~~~ & \multicolumn{3}{c}{Custom periodic} & ~~~~ & \multicolumn{3}{c}{Standard periodic}\tabularnewline
\cline{5-7} \cline{6-7} \cline{7-7} \cline{9-11} \cline{10-11} \cline{11-11} \cline{13-15} \cline{14-15} \cline{15-15} 
\multirow{1}{*}{System} & \multirow{1}{*}{Basis} & \multirow{1}{*}{Nbas} & ~~~ & Ovlp & Kin & Coul &  & Ovlp & Kin & Coul &  & Ovlp & Kin & Coul\tabularnewline
\cline{1-3} \cline{2-3} \cline{3-3} \cline{5-7} \cline{6-7} \cline{7-7} \cline{9-11} \cline{10-11} \cline{11-11} \cline{13-15} \cline{14-15} \cline{15-15} 
\textbf{Diamond} &  &  &  &  &  &  &  &  &  &  &  &  &  & \tabularnewline
 & univ-JKFIT & 600 &  & 0.02 & 0.02 & 0.01 &  & 0.14 & 0.15 & 0.19 &  & 5.73 & 6.45 & 5.32\tabularnewline
 & ANO-RCC & 728 &  & 0.02 & 0.02 & 0.02 &  & 0.15 & 0.20 & 0.24 &  & 21.73 & 24.94 & 9.85\tabularnewline
\textbf{Silicon} &  &  &  &  &  &  &  &  &  &  &  &  &  & \tabularnewline
 & univ-JKFIT & 1024 &  & 0.04 & 0.05 & 0.03 &  & 0.18 & 0.21 & 0.34 &  & 6.54 & 7.17 & 4.46\tabularnewline
 & ANO-RCC & 800 &  & 0.02 & 0.03 & 0.02 &  & 0.13 & 0.17 & 0.28 &  & 9.55 & 11.17 & 5.22\tabularnewline
\textbf{Ir$_{4}$Ba$_{4}$O$_{10}$} &  &  &  &  &  &  &  &  &  &  &  &  &  & \tabularnewline
 & univ-JKFIT & 3942 &  & 0.58 & 0.73 & 0.58 &  & 1.93 & 1.83 & 4.60 &  & 78.58 & 81.60 & 65.01\tabularnewline
 & ANO-RCC & 3870 &  & 0.52 & 0.70 & 0.73 &  & 1.01 & 4.11 & 7.81 &  & 576.55 & 590.31 & 316.89\tabularnewline
\hline 
\end{tabular}\caption{The table shows the cost of calculating the 2-center integrals with
overlap, kinetic, and Coulomb kernels in molecules (using PySCF) and
in periodic systems using the algorithm outlined in this paper (Custom
periodic) and the standard algorithm currently implemented in PySCF
(Standard periodic). For all kernels, the cost of calculating the
integrals in periodic systems using the custom algorithm is between
a factor of 5 and 10 times more expensive than in molecules. This
is a significant improvement over the standard PySCF algorithm.}
\label{tab:my_label-1}
\end{table}

The algorithm outlined in the current work is about a factor of 10
slower than molecular integral evaluation. This is to be contrasted
with the standard algorithm which is in some cases a factor of up
to 1000 slower than the molecular integral evaluation. In the standard
algorithm, the overlap and kinetic integrals are evaluated using summation
in the direct space and this becomes expensive when diffuse functions
are included in the basis set. It is interesting to note that in the
standard algorithm the coulomb integral evaluation is cheaper than
overlap and kinetic integrals. This is because, for coulomb kernel,
a background density is subtracted that makes the overall Gaussian
function charge-neutral \citep{Sun2017}. The contribution of the
background charge is then calculated in the reciprocal space. This
is similar in spirit to what we have proposed here, with the conceptual
difference that we split the kernel into the real and reciprocal space
and the practical difference that we use analytic Fourier transform
of the Gaussian basis functions instead of FFT. In the following,
we show that the present algorithm can be extended to calculate electron
densities that are needed in density functional theory calculations
and also for 3-center integrals which are used to approximate the
4-center coulomb integrals in the density fitting procedure \citep{whitten,dunlap}.

\section{Densities and 3-center integrals\label{sec:3-center}}

The approach outlined in this article can be extended to evaluate
electron densities and 3-electron integrals for periodic systems.
To evaluate the Gaussian densities one needs to be able to calculate
the value of a periodized Gaussian function at a grid point, 
\[
\rho_{a}\mathbf{\left(\mathbf{r}\right)}=\sum_{\mathbf{P}}\phi_{a}\left(\mathbf{r+P}\right),
\]
where $\phi_{a}$ is a Gaussian basis function and $\rho_{a}$ is
the contribution to density due to this basis function. The density
can be evaluated, as usual, by summation in real space for sharp Gaussians
and in reciprocal space for diffuse Gaussians by using the Poisson's
summation formula.

For molecules, the three center integrals of the $s-$type Gaussians
can be evaluated as follows \citep{ahlrichs06,ahlrichs04} 
\begin{align}
\left\langle ab|\frac{1}{r}|c\right\rangle =\frac{2\pi^{5/2}}{(a+b)c\sqrt{a+b+c}}S_{ab}G_{0}\left(\rho,T\right),
\end{align}
where $\alpha=\frac{ab}{a+b}$, $S_{ab}=\exp\left(-\frac{ab}{a+b}\left\Vert \mathbf{A}-\mathbf{B}\right\Vert {}^{2}\right)$,
$\rho=\frac{(a+b)c}{a+b+c}$ and $T=\rho\left\Vert \frac{a\mathbf{A}}{a+b}+\frac{b\mathbf{B}}{a+b}-\mathbf{C}\right\Vert {}^{2}$.

In the periodic case, the three center integrals acquire two additional
summations,
\begin{align}
\sum_{\mathbf{P,Q}}\left\langle ab_{\mathbf{Q}}|\frac{1}{r}|c_{\mathbf{P}}\right\rangle =\frac{2\pi^{5/2}}{(a+b)c\sqrt{a+b+c}}\sum_{\mathbf{P,Q}}\exp\left(-\frac{ab}{a+b}\left\Vert \mathbf{A}-\mathbf{B}-\mathbf{Q}\right\Vert {}^{2}\right)G_{0}\left(\rho,T\left(\mathbf{P,Q}\right)\right) & ,
\end{align}
or 
\begin{align}
\sum_{\mathbf{P,Q}}\left\langle ab_{\mathbf{Q}}|\frac{1}{r}|c_{\mathbf{P}}\right\rangle =\frac{2\pi^{5/2}}{(a+b)c\sqrt{a+b+c}}\sum_{\mathbf{Q}}\exp\left(-\frac{ab}{a+b}\left\Vert \mathbf{A}-\mathbf{B}-\mathbf{Q}\right\Vert {}^{2}\right)\sum_{\mathbf{P}}G_{0}\left(\rho,T\mathbf{\left(P,Q\right)}\right),
\end{align}
where $T\left(\mathbf{P,Q}\right)=\rho\left\Vert \frac{a\mathbf{A}}{a+b}+\frac{b\mathbf{B}}{a+b}-\mathbf{C}+\frac{b\mathbf{Q}}{a+b}-\mathbf{P}\right\Vert {}^{2}$.
We have already shown that the summation over $\mathbf{P}$ can be
treated efficiently by splitting it between the real and reciprocal
spaces. The inner summation over $\mathbf{P}$ can be split up into
three terms using Equation~\ref{eq:coulSum}; we then treat each
of these terms separately starting with the reciprocal summation.
We obtain
\begin{align}
 & =\frac{2\pi^{5/2}}{(a+b)c\sqrt{a+b+c}}\sum_{\mathbf{Q}}\exp(-\frac{ab}{a+b}\left\Vert \mathbf{A}-\mathbf{B}-\mathbf{Q}\right\Vert {}^{2})\sum_{\mathbf{G\neq0}}\frac{\exp\left(-\frac{G^{2}}{4\eta^{2}\rho}\right)}{G^{2}}\exp\left(-i\mathbf{G}\cdot\left(\frac{a\mathbf{A}}{a+b}+\frac{b\mathbf{B}}{a+b}-\mathbf{C}+\frac{b\mathbf{Q}}{a+b}\right)\right)\nonumber \\
 & =\frac{2\pi^{5/2}}{(a+b)c\sqrt{a+b+c}}\sum_{\mathbf{G\neq0}}\frac{\exp\left(-\frac{G^{2}}{4\eta^{2}\rho}\right)}{G^{2}}\left(\sum_{\mathbf{Q}}\exp(-\frac{ab}{a+b}\left\Vert \mathbf{A}-\mathbf{B}-\mathbf{Q}\right\Vert {}^{2})\exp\left(-i\mathbf{G}\cdot\left(\frac{a\mathbf{A}}{a+b}+\frac{b\mathbf{B}}{a+b}-\mathbf{C}+\frac{b\mathbf{Q}}{a+b}\right)\right)\right).\label{eq:FR}
\end{align}
In the second line we have rearranged terms so that the inner summation
can be readily converted to summation in Fourier space using the Poisson's
summation formula. The real space contribution 
\begin{align}
 & \sqrt{\frac{4(a+b)c}{\pi^{3}(a+b+c)}}\frac{2\pi^{5/2}}{(a+b)c\sqrt{a+b+c}}\sum_{\mathbf{Q}}\exp\left(-\frac{ab}{a+b}\left\Vert \mathbf{A}-\mathbf{B}-\mathbf{Q}\right\Vert {}^{2}\right)\sum_{\mathbf{P}}\left(F_{0}\left(T\mathbf{\left(P,Q\right)}\right)-\eta F_{0}\left(\eta^{2}T\left(\mathbf{P,Q}\right)\right)\right)\label{eq:RR}
\end{align}
is also evaluated efficiently by choosing $\eta$ such that the difference
$F_{0}\left(T\left(\mathbf{P,Q}\right)\right)-\eta F_{0}\left(\eta^{2}T\left(\mathbf{P,Q}\right)\right)$
is sufficiently small. By making use of the property that $F_{0}(T)\approx\sqrt{\frac{\pi}{4T}}$
for $T>30$ (to better than 14 decimal places), we can ensure that
both the $\mathbf{P}$ and $\mathbf{Q}$ summations converge rapidly
even when $\frac{ab}{\alpha}$ is small.

Work is currently under way for implementing these integrals.

\section{Conclusions}

In this article we have outlined an algorithm to efficiently calculate
2-center integrals for periodic systems. The algorithm partitions
the lattice summation between real and reciprocal space by making
use of the Poisson's summation formula. The real space summation can
be performed with any of the standard algorithms using the molecular
codes. We provide details on how the reciprocal space summation can
be performed efficiently. The technique is quite general and we have
shown that it can be used to with overlap, kinetic and Coulomb kernels.
Further, we have outlined an algorithm for calculating electron densities
and 3-center integrals.

\section{Acknowledgements}

SS was supported by NSF through the grant CHE-1800584. SS was also
partly supported through the Sloan research fellowship.


\end{document}